\documentclass[a4paper]{article}
\usepackage{amsmath}
\usepackage{graphicx}
\usepackage{geometry}
\usepackage{floatrow}
\usepackage{layout}
\usepackage{amssymb} 
\usepackage{multirow}
\usepackage{caption}
\usepackage{authblk}
\usepackage{hyperref}
 \usepackage{indentfirst}
\usepackage{physics} 
\usepackage{algorithm}
\usepackage{algorithmic}
\usepackage{subcaption}

\newcommand*{\email}[1]{%
    \normalsize\href{mailto:#1}{#1}\par
    }

\title{\textbf{\huge{Flow Annealed Kalman Inversion for Gradient-Free Inference in Bayesian Inverse Problems}}}
\author{Richard D.P. Grumitt $^{1}$}
\author{Minas Karamanis $^{2, 3}$}
\author{Uro\v{s} Seljak $^{2,3}$}

\affil{$^{1}$ \quad Department of Astronomy, Tsinghua University, Beijing 100084, China\\
$^{2}$ \quad Berkeley Center for Cosmological Physics and Department of Physics, University of California, Berkeley, CA 94720\\
$^{3}$ \quad Physics Department, Lawrence Berkeley National Laboratory, Cyclotron Rd, Berkeley, CA
94720\\
\email{rgrumitt@mail.tsinghua.edu.cn}}

\date{\today}

\begin{document}
\maketitle



\abstract{For many scientific inverse problems we are required to evaluate an expensive forward model. Moreover, the model is often given in such a form that it is unrealistic to access its gradients. In such a scenario, standard Markov Chain Monte Carlo algorithms quickly become impractical, requiring a large number of serial model evaluations to converge on the target distribution. In this paper we introduce Flow Annealed Kalman Inversion (FAKI). This is a generalization of Ensemble Kalman Inversion (EKI), where we embed the Kalman filter updates in a temperature annealing scheme, and use normalizing flows (NF) to map the intermediate measures corresponding to each temperature level to the standard Gaussian. In doing so, we relax the Gaussian ansatz for the intermediate measures used in standard EKI, allowing us to achieve higher fidelity approximations to non-Gaussian targets. We demonstrate the performance of FAKI on two numerical benchmarks, showing dramatic improvements over standard EKI in terms of accuracy whilst accelerating its already rapid convergence properties (typically in $\mathcal{O}(10)$ steps).}

\section{Introduction}

Many scientific inference tasks are concerned with inverse problems of the form
\begin{equation}
    y = \mathcal{G}(x)+\eta,
\end{equation}
where $y\in\mathbb{R}^{d_y}$ are the data, $x\in\mathbb{R}^{d}$ are the model parameters, $\mathcal{G}$ is the forward map, and $\eta$ is the observation noise. Throughout this work we will assume that we do not have access to gradients of $\mathcal{G}$ with respect to the parameters, and that $\eta\sim\mathcal{N}(0, \Gamma)$ where $\Gamma$ is a fixed noise covariance. The assumption of additive Gaussian noise is the standard setting for Ensemble Kalman Inversion (EKI) \cite{iglesias2013ensemble, iglesias2016regularizing, schillings2017analysis, chada2018parameterizations, schillings2018convergence, iglesias2021adaptive, huang2022iterated, huang2022efficient}, and whilst we are restricted to problems with Gaussian likelihoods, this covers a large family of scientific inverse problems. The goal of the Bayesian inverse problem is then to recover the posterior distribution over the model parameters given our observations, $p(x|y)$.

Typical gradient-free inference methods often involve some variant on Markov Chain Monte Carlo (MCMC) algorithms e.g., random walk Metropolis \cite{geyer1992MCMC, gelman1997RWM, cotter2013pCN}, or Sequential Monte Carlo (SMC) \cite{moral2006SMC}. However, these methods typically require $\gtrsim 10^3$ serial model evaluations to achieve convergence, making them intractable for problems with expensive forward models. EKI by contrast utilizes embarrassingly parallel model evaluations to update parameter estimates, typically converging to an approximate solution in $\mathcal{O}(10)$ iterations \cite{iglesias2016regularizing, iglesias2021adaptive, huang2022iterated, huang2022efficient}. 

EKI leverages ideas originally developed in the context of Ensemble Kalman Filtering (EKF) for data assimilation \cite{evensen1994sequential}. Since its development, EKI has seen applications across a range of disciplines, including studies of fluid flow \cite{xiao2016quantifying}, climate models \cite{schneider2017earth} and machine learning tasks \cite{kovachki2019ensemble}. EKI can be understood in the context of annealing, where seek to move from the prior to the posterior through a sequence of intermediate measures. In standard EKI, this involves constructing a sequence of Gaussian approximations to the intermediate measures. In the regime where we have a Gaussian prior $\pi_0(x)=\mathcal{N}(m_0, C_0)$ and a linear forward model $\mathcal{G}(x)=Gx$, the particle distribution obtained via EKI converges to the true posterior in the limit where the ensemble size $J\rightarrow\infty$. However, outside this linear, Gaussian regime EKI is an uncontrolled approximation to the posterior that is constructed on the basis of matching first and second moments of the target distribution. Nonetheless, EKI has been shown to perform well on problems with nonlinear forward models and slightly non-Gaussian targets \cite{iglesias2013ensemble, iglesias2016regularizing, iglesias2021adaptive}.

In this work we propose the application of normalizing flows (NF) \cite{dinh2016density, papamakarios2017masked, kingma2018glow, dai2021sliced} to relax the Gaussian ansatz made by standard EKI for the intermediate measures. Instead of assuming a Gaussian particle distribution at each iteration, the NF is used to fit for the empirical particle distribution and map to a Gaussian latent space, where the EKI updates are performed. In doing so, we are better able to capture non-Gaussian target geometries. The structure of this paper is as follows: in Section \ref{sec: methods} we describe the Flow Annealed Kalman Inversion (FAKI) algorithm, in Section \ref{sec: results} we demonstrate the performance of the method on two Bayesian inference tasks with non-Gaussian target geometries and we summarize our work in Section \ref{sec: conclusions}.  

\section{Methods}\label{sec: methods}

\subsection{Regularized Ensemble Kalman Inversion}\label{subsec: EKI}

A number of versions of EKI have been proposed in the literature. Of interest here is the regularized, perturbed observation form of EKI \cite{iglesias2021adaptive}. Starting with an ensemble of particles drawn from the prior, $\{x_0^j\}_{j=1}^{J}$, the particles are updated at each iteration according to
\begin{equation}
    x_{n+1}^j=x_{n}^j+C_n^{x\mathcal{G}}(C_n^{\mathcal{G}\mathcal{G}}+\alpha_n\Gamma)^{-1}(y-\mathcal{G}(x_n^j)+\sqrt{\alpha_n}\xi_n^j).
    \label{eqn: eki update}
\end{equation}
The empirical covariances $C_n^{x\mathcal{G}}$ and $C_n^{\mathcal{G}\mathcal{G}}$ are given by
\begin{align}
    C_n^{x\mathcal{G}} &= \frac{1}{J-1}\sum_{j=1}^J (x_n^j-\langle x_n\rangle)\otimes(\mathcal{G}(x_n^j)-\langle\mathcal{G}_n\rangle),\\
    C_n^{\mathcal{G}\mathcal{G}} &= \frac{1}{J-1}\sum_{j=1}^J (\mathcal{G}(x_n^j)-\langle\mathcal{G}_n\rangle)\otimes(\mathcal{G}(x_n^j)-\langle\mathcal{G}_n\rangle).
\end{align}
At each iteration we perturb the forward model evaluations with the Gaussian observation noise $\xi^j_n\sim\mathcal{N}(0, \Gamma)$. The parameter $\alpha_n$ is a Tikhonov regularisation parameter, which can be viewed as an inverse step size in the Bayesian annealing context. In particular, given a set of annealing parameters $\beta_0\equiv 0<\beta_1<\ldots<\beta_{N}<\beta_{N+1}\equiv 1$, we have the corresponding set of target distributions
\begin{equation}
    \pi_n(x)\propto\pi_0(x)\exp\left(-\frac{\beta_n}{2}\norm{\Gamma^{-1/2}(y-\mathcal{G}(x))}^2\right),
\end{equation}
with
\begin{equation}
    \alpha_n=\beta_{n+1}-\beta_n.
\end{equation}
EKI proceeds by constructing a sequence of ensemble approximations to Gaussian distributions that approximate the intermediate targets. 

The choice of the regularization parameter, $\alpha_n$ controls the transition from the prior to the posterior. Previous proposals for an adaptive choice have taken inspiration from SMC by using a threshold on the effective sample size (ESS) of the particles at each temeperature level \cite{de2018quantifying, iglesias2018bayesian}. In this work we adopt the same approach, calculating pseudo-importance weights at each temperature given by
\begin{equation}
    w_n^j=\exp\left( -\frac{1}{2}(\beta_{n+1} - \beta_n)\norm{\Gamma^{-1/2}(y-\mathcal{G}(x_n^j))}^2 \right).
\end{equation}
The next temperature level can then be selected by solving
\begin{equation}
    \left(\sum_{j=1}^J w_n^j(\beta_{n+1})^2\right)^{-1}\left(\sum_{j=1}^J w_n^j(\beta_{n+1})\right)^2 = \tau J,
    \label{eqn: bisection}
\end{equation}
using the bisection method, where $0<\tau<1$ is the target fractional ESS threshold. Throughout our work we set $\tau=0.5$. Full pseudocode for EKI is given in Algorithm \ref{alg: eki}. 

\begin{algorithm}
   \caption{Ensemble Kalman Inversion}
\begin{algorithmic}[1]
   \STATE {\bfseries Input:} $J$ prior samples $\{x_0^j\sim\pi_0(x)\}_{j=1}^J$, data $y$, observation error covariance $\Gamma$ and fractional ESS target threshold $\tau$
   \STATE Initialize inverse temperature $\beta_0=0$, iteration counter $n=0$
   \WHILE{$\beta<1$ \do}
        \STATE Evaluate $\mathcal{G}_j=\mathcal{G}(x_n^j)$, $j\in\{1,\ldots, J\}$
        \STATE Solve for $\beta_{n+1}$ using the bisection method with Equation \ref{eqn: bisection}
        \STATE $\alpha_n \leftarrow \beta_{n+1} - \beta_n$
        \STATE Update particles using Equation \ref{eqn: eki update} to obtain $\{x_{n+1}^j\}_{j=1}^J$
        \STATE $n \leftarrow n+1$
    \ENDWHILE
   \STATE {\bfseries Output}: Converged particle ensemble $\{x_{N}^j\}_{j=1}^J$
\end{algorithmic}
\label{alg: eki}
\end{algorithm}

\subsection{Normalizing Flows}

As discussed above, standard EKI proceeds by constructing a sequence of ensemble approximations to Gaussian distributions. The procedure works well in the situation where the target and all the intermediate measures are close to Gaussian. However, when any of these measures are far from Gaussian, EKI can dramatically fail to capture the final target geometry.

To address this shortcoming we propose the use of NFs to approximate each intermediate target, instead of using the Gaussian ansatz of standard EKI. NFs are powerful generative models that can be used for flexible density estimation and sampling \cite{dinh2016density, papamakarios2017masked, kingma2018glow, dai2021sliced}. An NF model maps from the original space $x\in\mathbb{R}^d$ to a latent space $z\in\mathbb{R}^d$, through a sequence of invertible transformations $f=f_1\circ f_2\circ\ldots\circ f_L$, such that we have a bijective mapping $z=f(x)$. The mapping is such that the latent variables are mapped to some simple base distribution, typically chosen to be the standard Normal distribution, giving $z\sim p_z(z)=\mathcal{N}(0, I)$.

The NF density can be evaluated through the change of variables formula,
\begin{equation}
    q(x) = p_z(f(x))\left|\mathrm{det}\,Df(x)\right| = p_z(f(x))\prod_{l=1}^L \left|\mathrm{det}\,Df_l(x)\right|,
\end{equation}
where $Df(x)=\partial f(x)/\partial x$ denotes the Jacobian of $f$. Efficient evaluation of this density requires the Jacobian of the transformation to be easy to evaluate, and efficient sampling requires the inverse of the mapping $f$ to be easy to calculate. In this work we use Masked Autoregressive Flows (MAF) \cite{papamakarios2017masked}, which have previously been found to perform well in the context of preconditioned MCMC sampling within SMC without the need for expensive hyper-parameter searches during sampling \cite{karamanis2022accelerating}.

\subsection{Flow Annealed Kalman Inversion}

Given particles distributed as $\pi_n(x)$, the subsequent target can be written as
\begin{equation}
    \pi_{n+1}(x)\propto \pi_n(x)\exp\left(-\frac{1}{2\alpha_n}\norm{\Gamma^{-1/2}(y-\mathcal{G}(x))}^2\right).
\end{equation}
We may therefore view $\pi_n(x)$ i.e., the posterior at the temperature level $\beta_n$, as an effective prior for $\pi_{n+1}(x)$, with a data likelihood annealed by $\alpha_{n}^{-1}$. By fitting an NF to the particles $\{x_n^j\}_{j=1}^J$, we obtain an approximate map from the intermediate target $\pi_n(x)$ to $\mathcal{N}(z|0, I)$. The latent space target is then given by the change of variables formula as
\begin{equation}
    \pi_{n+1}(z)=\pi_{n+1}(x=f_n^{-1}(z))\left|\mathrm{det}\,Df_n^{-1}(z)\right|.
\end{equation}
By controlling the choice of $\alpha_n$, we control the distance between the Gaussianized effective prior and this latent space target density. For FAKI, we therefore perform the EKI updates in the NF latent space at each temperature level, allowing us to relax the Gaussian ansatz of standard EKI by constructing an approximate map from each $\pi_n(x)$ to a Gaussian latent space. It is worth noting that, whilst this method relaxes the Gaussianity assumptions of standard EKI, it does not address the linearity assumptions used in deriving EKI.

The FAKI update for the latent space particle locations is given by  
\begin{equation}
    z_{n+1}^j = z_{n}^j+\mathcal{C}_n^{z\mathcal{G}}(\mathcal{C}_n^{\mathcal{G}\mathcal{G}}+\alpha_n\Gamma)^{-1}(y-\mathcal{G}(f_n^{-1}(z_n^j))+\sqrt{\alpha_n}\xi_n^j),
    \label{eqn: latent EKI}
\end{equation}
where the latent space empirical covariances are given by
\begin{align}
    \mathcal{C}_n^{z\mathcal{G}} &= \frac{1}{J-1}\sum_{j=1}^J (z_n^j-\langle z_n\rangle)\otimes(\mathcal{G}(f_n^{-1}(z_n^j))-\langle\mathcal{G}_n\rangle),\\
    \mathcal{C}_n^{\mathcal{G}\mathcal{G}} &= \frac{1}{J-1}\sum_{j=1}^J (\mathcal{G}(f_n^{-1}(z_n^j))-\langle\mathcal{G}_n\rangle)\otimes(\mathcal{G}(f_n^{-1}(z_n^j))-\langle\mathcal{G}_n\rangle).
    \label{eqn: latent cov}
\end{align}
Full pseudocode for FAKI is given in Algorithm \ref{alg: faki}.

\begin{algorithm}
   \caption{Flow Annealed Kalman Inversion}
\begin{algorithmic}[1]
   \STATE {\bfseries Input:} $J$ prior samples $\{x_0^j\sim\pi_0(x)\}_{j=1}^J$, data $y$, observation error covariance $\Gamma$ and fractional ESS target threshold $\tau$
   \STATE Initialize inverse temperature $\beta_0=0$, iteration counter $n=0$
   \WHILE{$\beta<1$ \do}
        \STATE Evaluate $\mathcal{G}_j=\mathcal{G}(x_n^j)$, $j\in\{1,\ldots, J\}$
        \STATE Solve for $\beta_{n+1}$ using the bisection method with Equation \ref{eqn: bisection}
        \STATE $\alpha_n \leftarrow \beta_{n+1} - \beta_n$
        \STATE Fit NF map $f_n$ to current samples $\{x_n^j\}_{j=1}^J$
        \STATE Map particles to latent space $z_n^j = f_n(x_n^j)$, $j\in\{1,\ldots, J\}$
        \STATE Update particles using Equation \ref{eqn: latent EKI} to obtain $\{z_{n+1}^j\}_{j=1}^J$
        \STATE Map back to the data space $x_{n+1}^j = f_n^{-1}(z_{n+1}^j)$, $j\in\{1,\ldots, J\}$
        \STATE $n \leftarrow n+1$
    \ENDWHILE
   \STATE {\bfseries Output}: Converged particle ensemble $\{x_{N}^j\}_{j=1}^J$
\end{algorithmic}
\label{alg: faki}
\end{algorithm}

\section{Results}\label{sec: results}

In this section we demonstrate the performance of FAKI compared to standard EKI on two numerical benchmarks, a two dimensional Rosenbrock distribution and a stochastic Lorenz system \cite{lorenz1963, ambrogioni2021automatic}. Both models display significant non-Gaussianity at some point during the transition from prior to posterior, severely frustrating the performance of EKI. This manifests in both reduced fidelity of the final ensemble approximations to the posterior, and in a larger number of iterations being required for convergence following the ESS-based annealing scheme described in Section \ref{subsec: EKI}.

In Table \ref{tab: numerical results} we provide statistics summarizing the performance of EKI and FAKI on our numerical benchmarks. We measure the quality of the posterior approximations by computing the 1-Wasserstein distance, $W_1$ \cite{villani_OT_2008, zhang2022pathfinder} between the samples obtained through FAKI and EKI, against reference posterior samples obtained via long runs of Hamiltonian Monte Carlo (HMC) \cite{neal2011mcmc, hoffman2014no}. These reference samples are thinned to be approximately independent when computing the 1-Wasserstein distances\footnote{We use the Python Wasserstein library: https://github.com/pkomiske/Wasserstein/.}. The 1-Wasserstein distance may be interpreted as the cost involved in rearranging one probability measure to look like another, with lower values indicating the two probability measures are closer to one another. In addition to this assessment of the approximation quality, we report the number of iterations, $N_\mathrm{iter}$ required by FAKI and EKI for convergence. For both quantities we report the median and median absolute deviation (MAD), estimated over 10 independent runs using different random seeds.

\begin{table}[H] 
\caption{Median and MAD values for $N_\mathrm{iter}$ and 1-Wasserstein distances for each model and algorithm combination, calculated over 10 independent runs using different random seeds. For both the numerical benchmarks we see that FAKI results in a reduced number of iterations for convergence, and a lower value of the 1-Wasserstein distance between the converged samples and the ground truth.\label{tab: numerical results}}
\begin{tabular}[\textwidth]{|c|c|c|c|c|c|c|}
\hline
Model &  Algorithm & $\mathrm{Median}[N_\mathrm{iter}]$	& $\mathrm{MAD}[N_\mathrm{iter}]$ & $\mathrm{Median}[W_1]$ & $\mathrm{MAD}[W_1]$\\
\hline
Rosenbrock & EKI &  100 & 7.0 & 0.72 & 0.05\\
Rosenbrock & FAKI &  34.0 & 7.0 & 0.43 & 0.14\\
Lorenz & EKI & 10.0 & 0.0 & 69.8 & 1.08\\
Lorenz & FAKI & 8.0 & 0.0 & 5.65 & 0.86\\
\hline
\end{tabular}
\end{table}

\subsection{$d=2$ Rosenbrock}

In our first numerical experiment we consider the two dimensional Rosenbrock distribution. This toy model allows us to clearly see the impact of non-Gaussianity on the performance of EKI, and how FAKI is able to alleviate these issues. For the Rosenbrock model we assume a Gaussian prior over the parameters $x\in\mathbb{R}^2$,
\begin{equation}
    x\sim\mathcal{N}(0, 10^2I).
\end{equation}
The data, $y\in\mathbb{R}^2$ are distributed according to the likelihood,
\begin{equation}
    y\sim\mathcal{N}(\mathcal{G}(x)=(x_1-x_0^2, x_0)^\intercal, \Gamma=\mathrm{diag}(0.01^2, 1^2)).
\end{equation}
To generate simulated data we evaluate $y=G((1, 1)^\intercal)+\eta$, where $\eta\sim\mathcal{N}(0, \mathrm{diag}(0.01^2, 1^2))$. The large difference in noise scales results in a highly non-Gaussian posterior geometry that poses a significant challenge for EKI. For each run of EKI and FAKI we use 100 particles. 

\begin{figure}[H]
\centering
      \begin{subfigure}[H]{0.48\linewidth}
        \includegraphics[width=\textwidth]{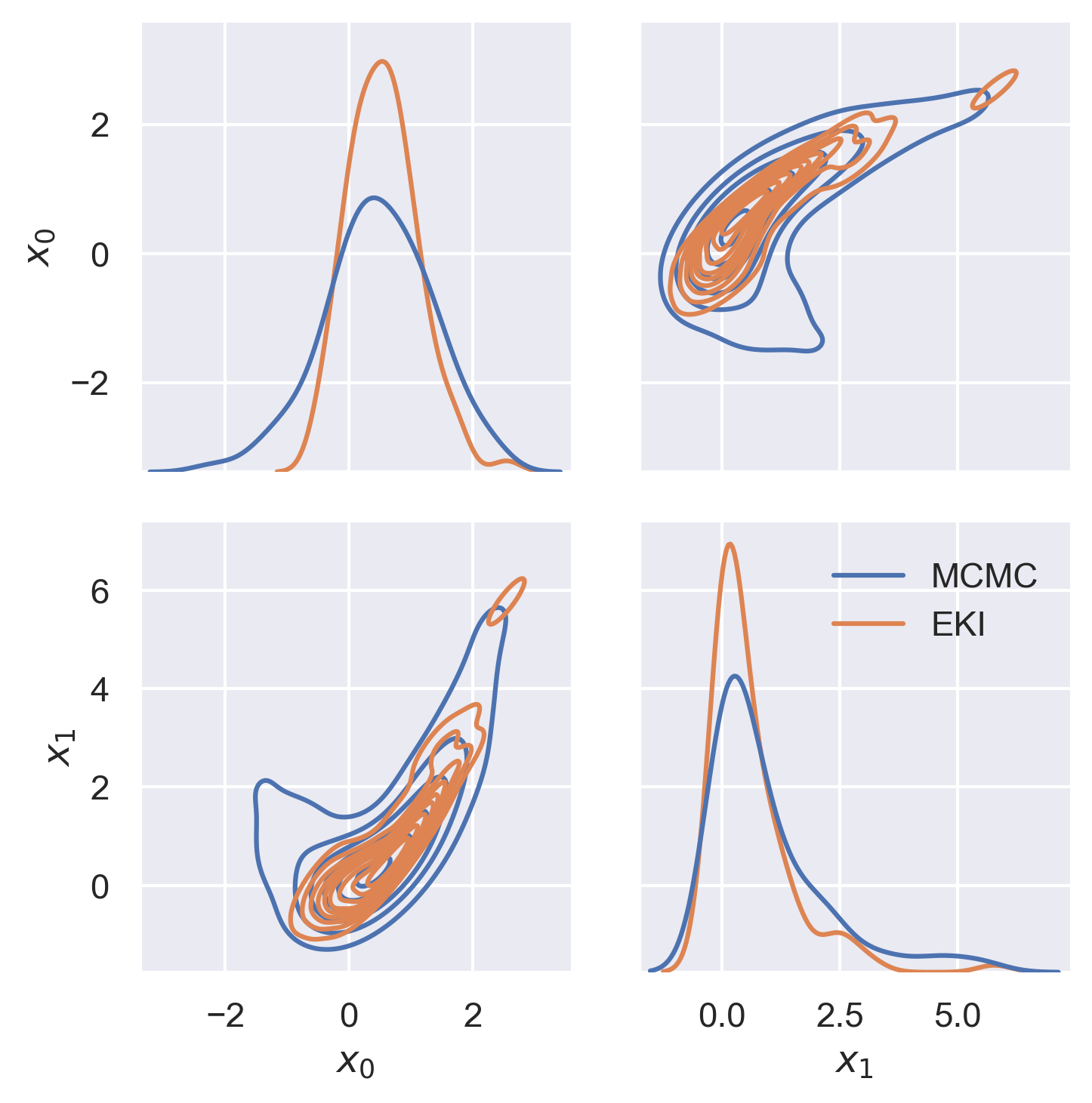}
        \caption{EKI-MCMC comparison}
      \end{subfigure}
      \begin{subfigure}[H]{0.48\linewidth}
        \includegraphics[width=\textwidth]{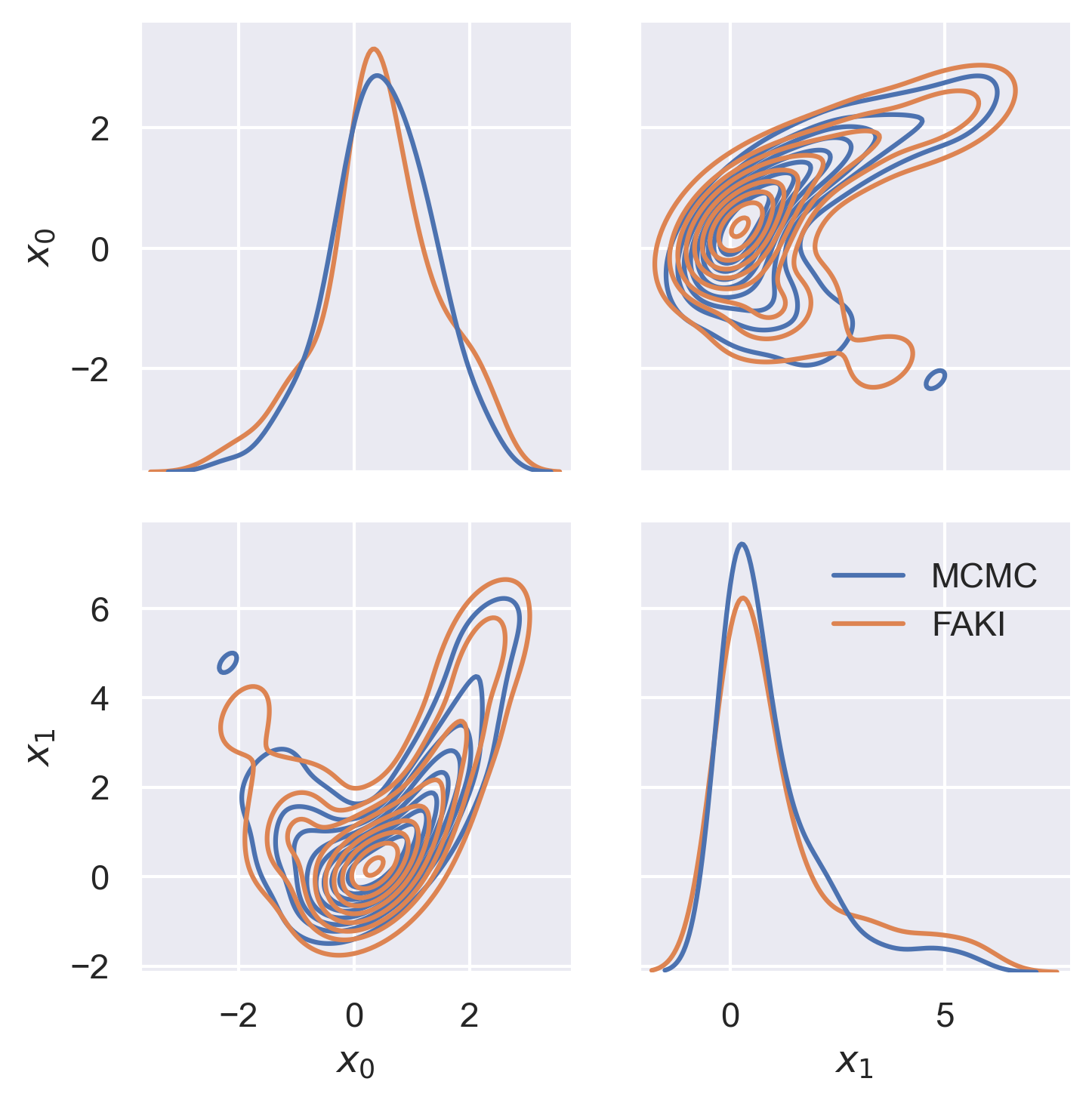}
        \caption{FAKI-MCMC comparison}
      \end{subfigure}
\caption{Pair-plots for the Rosenbrock target. Panel (a): Pair-plot comparison of samples from EKI and a long HMC run. Panel (b): Pair-plot comparison of samples from FAKI and a long HMC run. Samples from FAKI are able to correctly capture the highly nonlinear target geometry. Standard EKI struggles to fill the tails of the target, and requires $\sim 100$ iterations to converge, compared to $\sim 34$ iterations for FAKI.}
\label{fig: 2d rosenbrock scatter}
\end{figure}

In Figure~\ref{fig: 2d rosenbrock scatter} we show pair-plots comparing the final particle distributions obtained with EKI and FAKI against samples obtained through a long run of HMC. The NF mapping means that the ensemble approximation obtained by FAKI is able to capture the highly nonlinear target geometry. In comparison, EKI struggles to fill the tails of the Rosenbrock target. Moreover, whilst FAKI converges within $\sim 34$ iterations, EKI required a median number of $\sim 100$ iterations to converge using the ESS-based annealing scheme.  

\subsection{Stochastic Lorenz System}

The Lorenz equations are a set of coupled differential equations used as a simple model of atmospheric convection. Notably, for certain parameter values the Lorenz equations are known to exhibit chaotic behaviour \cite{lorenz1963}. In this work we follow \cite{ambrogioni2021automatic} and consider the stochastic Lorenz system,
\begin{align}
    \mathrm{d}X_t &= 10(Y_t - X_t)\mathrm{d}t + \mathrm{d}W^x_t, \label{eqn: x lorenz sde}\\
    \mathrm{d}Y_t &= X_t(28 - Z_t)\mathrm{d}t - Y_t\mathrm{d}t + \mathrm{d}W^y_t, \label{eqn: y lorenz sde}\\
    \mathrm{d}Z_t &= X_t Y_t\mathrm{d}t - \frac{8}{3}Z_t\mathrm{d}t + \mathrm{d}W^z_t,
    \label{eqn: z lorenz sde}
\end{align}
where $W^x_t$, $W^y_t$ and $W^z_t$ are Gaussian white noise processes with standard deviation $\sigma_0=0.1$. To generate simulated data we integrated these equations using an Euler–Maruyama scheme with $\mathrm{d}t=0.02$ for 30 steps, with initial conditions $X_0, Y_0, Z_0 \sim\mathcal{N}(0, 1^2)$. The observations are then taken to be the $X_t$ values over these 30 time steps, with Gaussian observational noise $\eta_t\sim\mathcal{N}(0, \sigma^2=1^2)$.

\begin{figure}[H]
\centering
      \begin{subfigure}[H]{0.48\linewidth}
        \includegraphics[width=\textwidth]{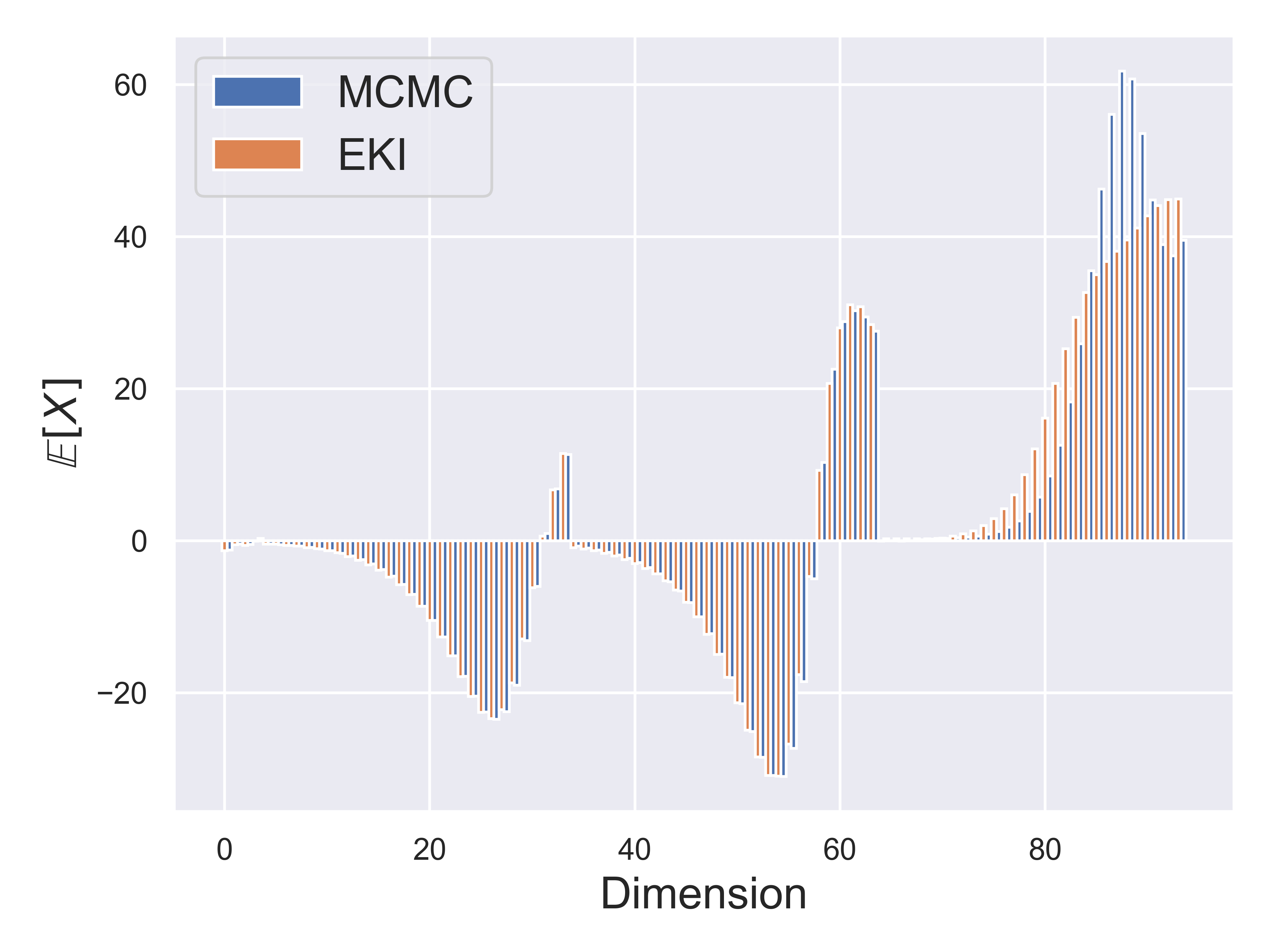}
        \caption{EKI-MCMC $\mathbb{E}[X]$ comparison}
      \end{subfigure}
      \begin{subfigure}[H]{0.48\linewidth}
        \includegraphics[width=\textwidth]{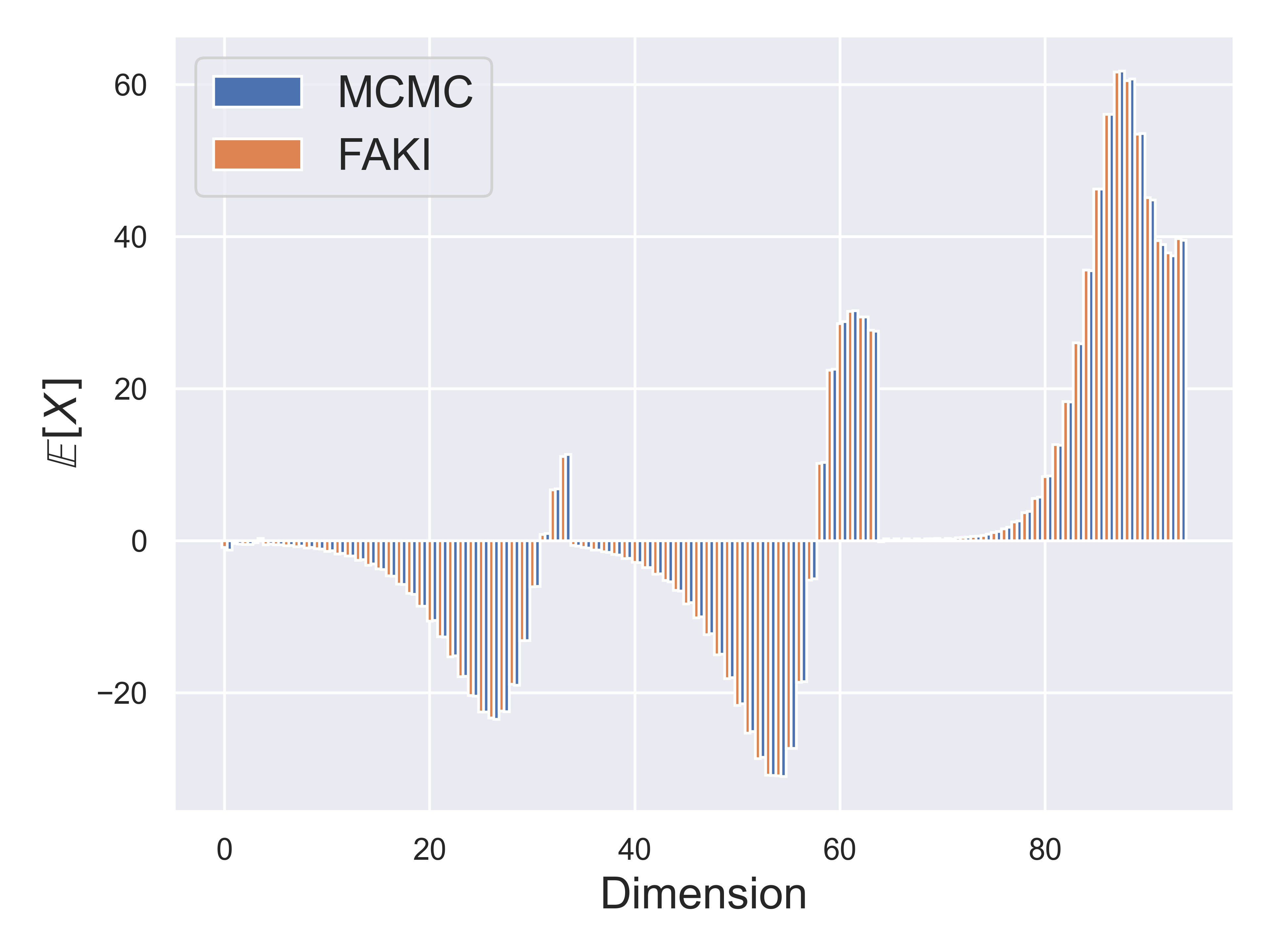}
        \caption{FAKI-MCMC $\mathbb{E}[X]$ comparison}
      \end{subfigure}
      \par\bigskip
      \begin{subfigure}[H]{0.48\linewidth}
        \includegraphics[width=\textwidth]{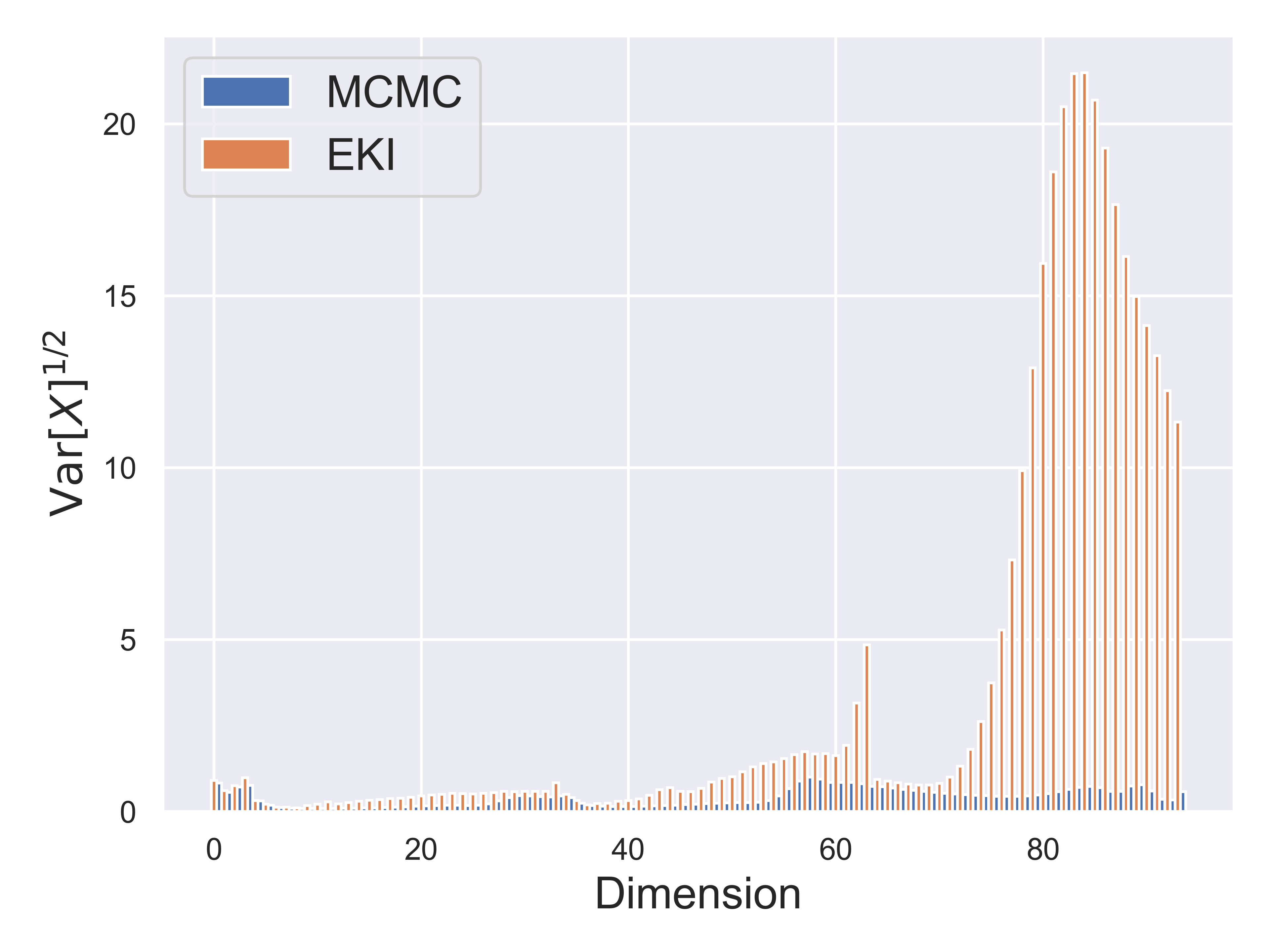}
        \caption{EKI-MCMC $\mathrm{Var}[X]^{1/2}$ comparison}
      \end{subfigure}
      \begin{subfigure}[H]{0.48\linewidth}
        \includegraphics[width=\textwidth]{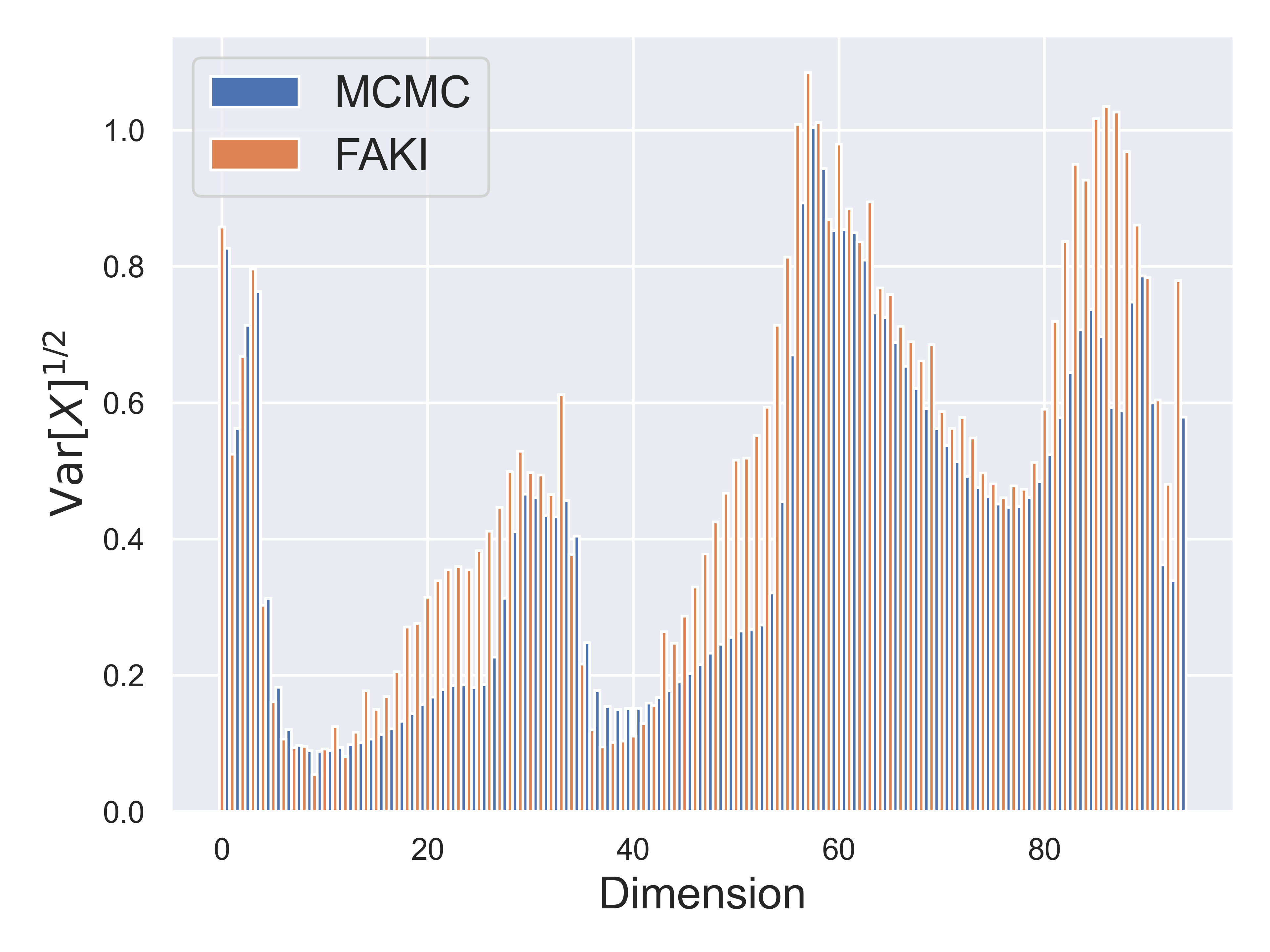}
        \caption{FAKI-MCMC $\mathrm{Var}[X]^{1/2}$ comparison}
      \end{subfigure}
\caption{Comparison of first and second moment estimates along each dimension for the stochastic Lorenz system. Panel (a): Comparison between the mean estimates from EKI and a long HMC run. Panel (b): Comparison between the mean estimates from FAKI and a long HMC run. Panel (c): Comparison between the standard deviation estimates from EKI and a long HMC run. Panel (d): Comparison between the standard deviation estimates from FAKI and a long HMC run. Blue bars indicate the moment estimates obtained via HMC along each dimension, with the adjacent orange bars showing the estimates obtained through EKI/FAKI. EKI is unable to obtain accurate mean estimates for much of the $Z_t$ trajectory, whilst FAKI is able to obtain accurate mean estimates for each dimension. FAKI outperforms EKI in its estimates of the marginal standard deviations, with EKI drastically overestimating the standard deviations along many dimensions.}
\label{fig: lorenz moments}
\end{figure} 

The goal of our inference here is to recover the initial conditions, the trajectories $(X_t, Y_t, Z_t)$ and the innovation noise scale $\sigma_0$, giving a parameter space of $d=94$ dimensions. We assign priors over these parameters as,
\begin{align}
    \log\sigma_0 &\sim \mathcal{N}(-1, 1^2),\\
    X_0, Y_0, Z_0 &\sim \mathcal{N}(0, 1^2),\\
    X_t &\sim \mathcal{N}(X_{t-1} + f_X(X_{t-1}, Y_{t-1}, Z_{t-1}, t-1)\mathrm{d}t, \sigma_0^2\mathrm{d}t), t\in\{1,\ldots, 30\},\\
    Y_t &\sim \mathcal{N}(Y_{t-1} + f_Y(X_{t-1}, Y_{t-1}, Z_{t-1}, t-1)\mathrm{d}t, \sigma_0^2\mathrm{d}t), t\in\{1,\ldots, 30\},\\
    Z_t &\sim \mathcal{N}(Z_{t-1} + f_Z(X_{t-1}, Y_{t-1}, Z_{t-1}, t-1)\mathrm{d}t, \sigma_0^2\mathrm{d}t), t\in\{1,\ldots, 30\},
\end{align}
where $f_X$, $f_Y$ and $f_Z$ are the transition functions corresponding to Equations \ref{eqn: x lorenz sde}-\ref{eqn: z lorenz sde} respectively. The Gaussian likelihood has the form
\begin{equation}
    \hat{X}_t\sim\mathcal{N}(X_t, \sigma^2), t\in\{1,\ldots, 30\},
\end{equation}
where $\hat{X}_t$ are the observations of the $X_t$ trajectory. The chaotic dynamics of the Lorenz system results in a highly non-Gaussian prior distribution, with the inversion having to proceed through a sequence of highly non-Gaussian intermediate measures towards the posterior. This severely frustrates the performance of EKI, with the Gaussian ansatz failing to describe the geometry of the intermediate measures. For each run of EKI and FAKI we use 940 particles.

In Figure \ref{fig: lorenz moments} we show the ensemble estimates for the mean and standard deviation along each dimension obtained by EKI and FAKI, compared to reference estimates obtained through long runs of HMC. FAKI is able to obtain accurate mean estimates along each dimension, whereas EKI is unable to obtain the correct means for much of the $Z_t$ trajectory. EKI severely overestimates the marginal standard deviations along many dimensions. This situation is alleviated by the NF mappings learned by FAKI. The greater fidelity of the FAKI posterior approximations are reflected in the median estimates for the 1-Wasserstein distances, with a value of $5.65$ for FAKI and $69.8$ for EKI.

\section{Conclusions}\label{sec: conclusions}

In this work we have introduced Flow Annealed Kalman Inversion (FAKI), a gradient-free inference algorithm for Bayesian inverse problems with expensive forward models. This is a generalization of Ensemble Kalman Inversion (EKI), where we utilize Normalizing Flows (NF) to replace the Gaussian ansatz made in EKI. Instead of constructing a sequence of ensemble approximations to Gaussian measures that approximate a sequence of intermediate measures, as we move from the prior to the posterior, we learn an NF mapping at each iteration to a Gaussian latent space. Provided the transition between temperature levels is controlled, we can perform Kalman inversion updates in the NF latent space. In the NF latent space, the Gaussianity assumptions of EKI are more closely satisfied, resulting in a more stable inversion at each temperature level.

We demonstrate the performance of FAKI on two numerical benchmarks, a $d=2$ Rosenbrock distribution and a $d=94$ stochastic Lorenz system. Both examples exhibit significant non-Gaussianity in the transition from prior to posterior that frustrate standard EKI. In the presence of strong non-Gaussianity, we find FAKI produces higher fidelity posterior approximations compared to EKI, as measured by the 1-Wasserstein distance between FAKI/EKI samples and reference HMC samples. In addition to the improved fidelity of the posterior approximations, we find FAKI tends to reduce the number of iterations required for convergence.

Whilst the application of NFs is able to relax the Gaussian ansatz of EKI, it does not address the linearity assumptions used in deriving EKI. As such, FAKI is still not exact for general forward models. In future work, it will be interesting to explore methods to address this, for example the combination of FAKI with unbiased MCMC or importance sampling methods. It would also be interesting to consider generalizations of FAKI that are able to accommodate non-Gaussian likelihoods and/or parameter-dependent noise covariances. The use of NFs means that we typically require ensemble sizes $J\gtrsim 10d$ to learn accurate NF maps with the MAF architecture employed in this work. It would be useful to explore alternative NF architectures and regularization schemes that are able to learn accurate NF maps with smaller ensemble sizes, in order to enable FAKI to scale to higher dimensions. In this work, we have found that the MAF architecture is able to capture a wide range of target geometries without the need for expensive NF hyper-parameter searches. However, it may be possible to exploit NF architectures with inductive biases that are particularly suited to common target geometries e.g., the nonlinear correlations that often appear in hierarchical models.   

\vspace{6pt}

\section*{Acknowledgments}

This research was funded by NSFC (grant No. 12250410240) and the U.S. Department of Energy, Office of Science, Office of Advanced Scientific Computing Research under Contract No. DE-AC02-05CH11231 at Lawrence Berkeley National Laboratory to enable research for Data-intensive Machine Learning and Analysis. RDPG was supported by a Tsinghua Shui Mu Fellowship.

The authors thank Qijia Jiang and David Nabergoj for helpful discussions.

\bibliographystyle{unsrt}
\bibliography{refs}

\end{document}